\documentclass[twocolumn, 
  aps, pra,
  amsmath,amssymb,
  longbibliography,
  ]{revtex4-1}
\usepackage{graphicx,color}
\usepackage{amsmath}
\usepackage{natbib}
\usepackage{epsfig}
\begin{document}

\title{\color{blue} Vibrational model of heat transfer in strongly coupled Yukawa fluids (dusty plasma liquids)}

\author{Sergey A. Khrapak}
\email{Sergey.Khrapak@gmx.de}
\affiliation{Joint Institute for High Temperatures, Russian Academy of Sciences, 125412 Moscow, Russia 
}

\begin{abstract}
A concise overview of the vibrational model of heat transfer in simple fluids with soft pairwise interactions is presented. The model is applied to evaluate the thermal conductivity coefficient of the strongly coupled Yukawa fluid, which often serves as a simplest model of a real liquid-like dusty (complex) plasma. A reasonable agreement with the available data from molecular dynamics numerical simulations is observed. Universality of the properly reduced thermal conductivity coefficient with respect to the effective coupling parameter is examined. Relations between the vibrational model and the excess entropy scaling of the thermal conductivity coefficient are discussed. 
\end{abstract}

\date{\today}

\maketitle

\section{Introduction}

Thermal conductivity is an important transport property of gases, plasmas and condensed matter. In conventional condensed matter it can be investigated experimentally only on macroscopic scales, because the motion of individual atoms cannot be usually resolved. Complex (dusty) plasma represents a system where heat transport (as well as many other phenomena) can be studied at the most fundamental individual particle kinetic level.   

A complex plasma consists of an ionized gas and charged micron-size particles~\cite{TsytovichUFN1997,FortovUFN,FortovPR,FortovBook}. When the main charging mechanism is the collection of electrons and ions from the surrounding plasma, the particle charge saturates at large negative values, because electrons are much more mobile than ions. The strong mutual interaction between the highly charged charged particles forces them to organize in a liquid-like or even crystalline structures. The characteristic particle sizes ($\sim 1$ $\mu$m), interparticle separations ($\sim 100$ $\mu$m), and characteristic dynamical frequency scales ($\sim 10$ Hz) are such that the motion of individual particles can be observed in real time. Heat transfer through the particle component can therefore be studied at an atomistic level.

In this context complex plasma is considered as a useful model system, where particles imitate the dynamics of atoms in real condensed matter systems. Plasma electrons and ions are treated as a neutralizing background, which provides screening to the repulsive Coulomb interactions between the charged particles. Electron contribution to the heat transfer, which would dominate heat conduction in real multi-component plasma-related systems and liquid metals is not considered. The focus is on how the energy propagates through the strongly coupled particle component.

Significant experimental efforts have been undertaken to investigate heat transfer in crystalline, melted, and
liquid-like complex plasma under different conditions~\cite{NunomuraPRL2005,VaulinaAIPProc2005,FortovPPR2006,FortovPRE2007,
NosenkoPRL2008,GoreePPCF2013,DuPRE2014,NosenkoPoP2021}.
Related numerical simulations have been mostly focused on heat conduction in screened Coulomb (Yukawa) systems, which serve as simplest models of real complex plasmas~\cite{SalinPRL2002,SalinPoP2003,DonkoPRE2004,HouJPA2009,KhrustalyovPRE2012,
ScheinerPRE2019,KahlertPPR2020}. However, a simple theory capable to explain and predict the values and main trends exhibited by the thermal conductivity coefficient has still been missing. A recently proposed vibrational model of heat transfer in fluids with soft interparticle interactions~\cite{KhrapakPRE01_2021} seems to be able to fill this gap. The purpose of this paper is to provide a concise overview of the model, demonstrate its applicability to strongly coupled Yukawa fluids, and to discuss some general properties characterizing the heat conductivity coefficient in fluids.    

\section{Vibrational model of heat transfer}

The vibrational paradigm of atomic dynamics in dense liquids has been discussed by many authors over many decades, see e.g. Refs.~\cite{FrenkelBook,Hubbard1969,Stillinger1982,ZwanzigJCP1983}. It applies to both, plasma-related and conventional neutral liquids. The only specifics of plasma-related fluids is the softness of the interaction potential and the absence of the long range attraction if Yukawa interaction potential is adopted (see below). The main assumptions behind the vibrational paradigm are as follows~\cite{KhrapakMolecules12_2021}:
Atoms exhibit solid-like oscillations about temporary equilibrium positions corresponding to a local minimum on the system's potential energy surface~\cite{FrenkelBook,Stillinger1982}; These positions do not form a regular lattice like in crystalline solids, but correspond to a liquid-like arrangement of atoms~\cite{Hubbard1969}; They are also not fixed, and change (diffuse or drift) with time (this is why liquids can flow), but on much longer time scales.  Effectively, one can assume that local configurations of atoms are preserved for some time until a fluctuation in the kinetic energy allows rearranging the positions of some of these atoms towards a new local minimum in the multidimensional potential energy surface. This picture allows to make important approximations about the properties of atomic motion and mechanisms of momentum and energy transport in the liquid state. For instance, the Stokes-Einstein (SE) relation without the hydrodynamic radius naturally arises within this vibrational paradigm under few additional assumptions~\cite{ZwanzigJCP1983,KhrapakMolPhys2019,KhrapakPRE10_2021,
KhrapakMolecules12_2021}.  

For energy transfer, separation of time scales corresponding to fast solid-like atomic oscillations and their slow drift plays a dominant role. Namely, it is reasonable to assume that a vibrating atom transports energy from its hotter to its cooler neighbors with a characteristic energy exchange rate equal to its average vibrational frequency, $\nu=\langle \omega\rangle/2\pi$. Each atom controls the energy transfer through an area of order $\Delta^2$, where $\Delta=\rho^{-1/3}$ is the average interatomic separation. A liquid can be approximated by a quasi-layered structure
with quasi-layers perpendicular to the temperature gradient (applied along $x$-axis) and
separated by the distance $\Delta$ (see e.g. Fig.~1 from Ref.~\cite{KhrapakPRE01_2021}). The energy difference between two neighbouring layers is $(dU/dx)\Delta$. The energy flux density can be then approximated as
\begin{equation}\label{Flow}
j\simeq -\frac{\nu}{\Delta^2}\left(\frac{dU}{dx}\right)\Delta= -\frac{\langle \omega\rangle}{2\pi \Delta}c_{\rm p} \frac{dT}{dx},
\end{equation} 
where $c_{\rm p}=(dU/dT)_{\rm p}$ is specific heat at constant pressure and the minus sign implies that the heat flow is down the temperature gradient. On the other hand, Fourier's law for the heat flow reads  
\begin{equation}\label{Fourier}
j=-\lambda\frac{dT}{dx}.
\end{equation}
Comparing Eqs.~(\ref{Flow}) and (\ref{Fourier}) we immediately obtain
\begin{equation}\label{lambda}
\lambda=c_{\rm p}\frac{\langle \omega\rangle}{2\pi \Delta}.
\end{equation}
This is essentially the expression derived in Ref.~\cite{KhrapakPRE01_2021}, except that the specific heat at constant volume, $c_{\rm v}$, appeared there. The present choice with $c_{\rm p}$ instead of $c_{\rm v}$ seems in general more appropriate, because pressure should be constant in equilibrium. The difference is insignificant for soft spheres, because dense fluids can be considered as essentially incompressible in a wide portion of their phase diagram not too far from the freezing curve, and thus $c_{\rm p}\simeq c_{\rm v}$ holds. For example, Eq.~(\ref{lambda}) with $c_{\rm p}\rightarrow c_{\rm v}$ works very well for soft plasma-related Coulomb and Yukawa interactions and the Lennard-Jones fluid~\cite{KhrapakPRE01_2021,KhrapakPoP2021,KhrapakPoP08_2021}.     
In such cases it is more appropriate to use $c_{\rm v}$ for practical estimates (since $c_{\rm v}$ is normally easier to evaluate). We follow this approach in the present manuscript. 


Since the actual frequency distribution can be quite complex
in liquids and can vary from one type of liquid to another,
some simplifying assumptions have to be employed to evaluate $\langle \omega \rangle$ in Eq.~(\ref{lambda}). In the simplest Einstein picture all atoms vibrate with the same (Einstein) frequency $\Omega_{\rm E}$.
The results in 
\begin{equation}\label{Horrocks}
\lambda = c_{\rm v} \frac{\Omega_{\rm E}}{2\pi \Delta}.
\end{equation}
This is very close to the expression proposed by Horrocks and
McLaughlin~\cite{Horrocks1960} (neglecting some difference in numerical coefficients). Below we consider application of this expression to the strongly coupled Yukawa fluid.

\section{Results}

The pairwise Yukawa repulsive interaction  potential (also known as screened Coulomb or Debye-H\"uckel potential) is
\begin{equation}\label{Yukawa}
\phi(r)=(Q^2/r)\exp(-\kappa r/a),
\end{equation}
where $Q$ is the particle charge and $\kappa$ is the dimensionless screening parameter, which is the ratio of the Wigner-Seitz radius $a=(4\pi \rho/3)^{-1/3}$ to the plasma screening length. This potential represents a reasonable first approximation to describe the actual (often more sophisticated) interactions between the charged particles immersed in a plasma medium~\cite{TsytovichUFN1997,FortovUFN,FortovPR,FortovBook,
KhrapakPRL2008,KhrapakCPP2009,ChaudhuriSM2011,IvlevBook,LampePoP2015}. 

The dynamics and thermodynamics of Yukawa systems are conventionally characterized by the screening parameter $\kappa$ and the coupling parameter $\Gamma=Q^2/aT$, where 
$T$ is the system temperature (in energy units).
The coupling parameter characterizes the ratio between the potential energy of interparticle interaction and the kinetic energy. In strongly coupled Yukawa fluids the condition $\Gamma\gg 1$ should be satisfied. For even higher coupling Yukawa fluids crystallize, forming either the body-centered-cubic (bcc) or face-centered-cubic (fcc) lattices. Detailed phase diagrams of Yukawa systems are available in the literature~\cite{RobbinsJCP1988,HamaguchiJCP1996,HamaguchiPRE1997,
VaulinaJETP2000,VaulinaPRE2002}. 
The screening parameter $\kappa$ determines the softness of the interparticle repulsion. It varies from the extremely soft and long-ranged Coulomb potential at $\kappa\rightarrow 0$ (one-component plasma limit) to the hard-sphere-like interaction limit at $\kappa\rightarrow \infty$. In the context of complex plasmas and colloidal suspensions the relatively ``soft'' regime, $\kappa\sim {\mathcal O}(1)$, is of particular interest. Most of previous investigations have focused on this regime and we follow this trend.


As already mentioned, a significant ammount of simulation data regarding thermal conductivity of Yukawa systems has been published. For a detailed comparison with the vibrational model we chose the data from non-equilibrium molecular dynamics (MD) simulations reported by Donko and Hartmann~\cite{DonkoPRE2004}. This is the most complete set of data in terms of covered $\Gamma$ and $\kappa$ presently available. To simplify the comparison we employ the Rosenfeld's normalization~\cite{RosenfeldJPCM1999} throughout this paper. The reduced thermal conductivity coefficient is $\lambda_{\rm R}=\lambda \Delta^2/v_{\rm T}$ (where $v_{\rm T}=\sqrt{T/m}$ is the thermal velocity). This is different from the normalization used originally, $\lambda'=\lambda/(\rho\omega_{p}a^2)$, where $\omega_{\rm p}=\sqrt{4\pi Q^2 \rho/m}$ is the plasma frequency scale. Note that since we express the temperature in energy units, the Boltzmann constant does not show up in the expressions (effectively $k_{\rm B}=1$). 

\begin{figure}
\includegraphics[width=8.cm]{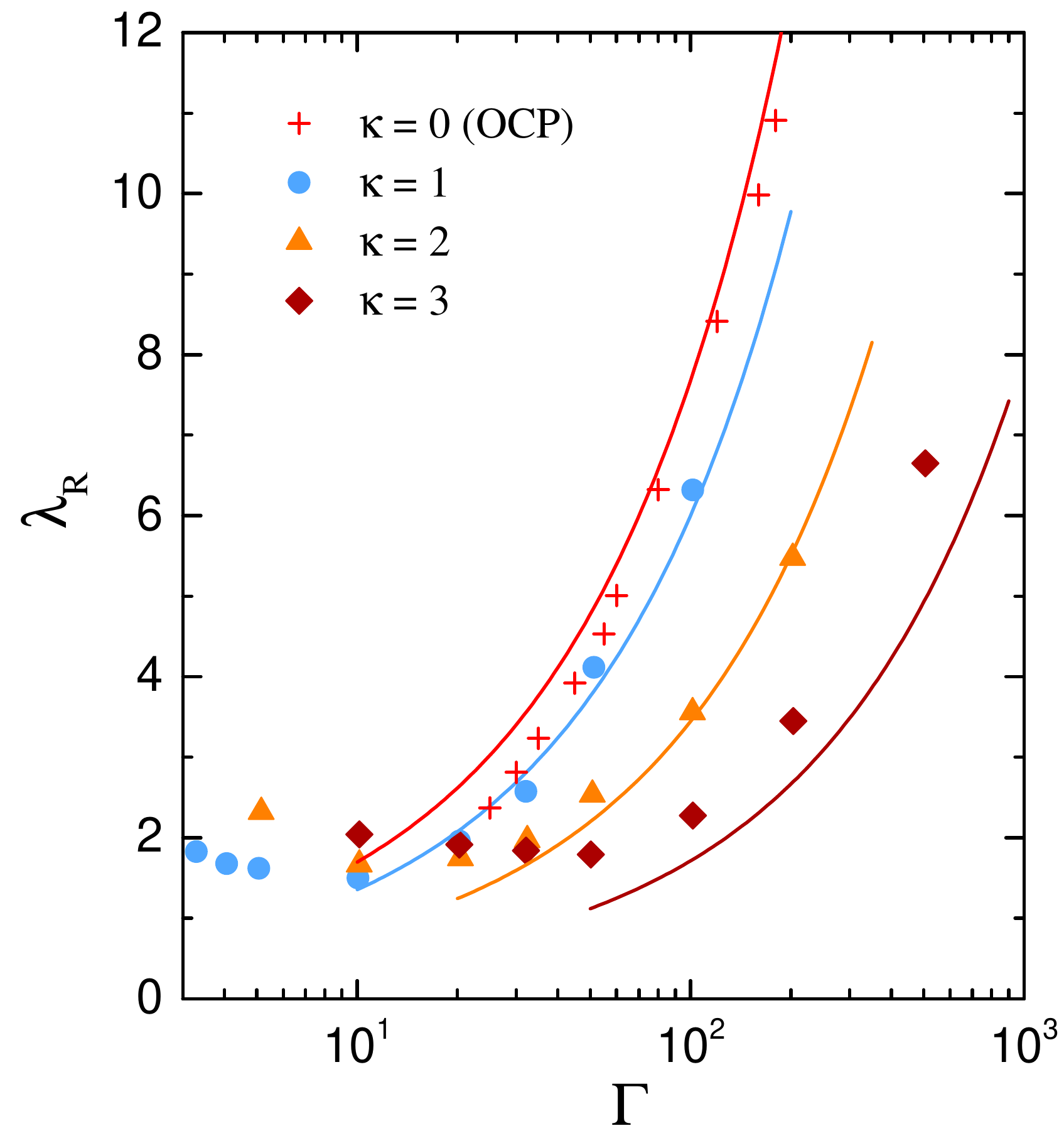}
\caption{(Color online) Reduced thermal conductivity coefficient $\lambda_{\rm R}$ versus the coupling parameter $\Gamma$. Citcles, triangles, and rhombs correspond to molecular dynamics results from Ref.~\cite{DonkoPRE2004} for $\kappa=1$, $\kappa=2$, and $\kappa=3$, respectively. Crosses are the OCP numerical results from Ref.~\cite{ScheinerPRE2019}. The solid curves are the theoretical calculations using Eqs.~(\ref{Horrocks}) and (\ref{cv}).}
\label{Fig1}
\end{figure}   
%
%

To evaluate the specific heat $c_{\rm v}$, the equation of state for the Yukawa fluid is required. Thermodynamics of Yukawa fluids has been relatively well investigated and understood~\cite{HamaguchiPRE1997,FaussurierPRE2003,ToliasPRE2014,
KhrapakPRE02_2015,KhrapakPPCF2015,
KhrapakJCP2015,ToliasPoP2015,ToliasPoP2019}. Various practical expressions for thermodynamic functions have been proposed in the literature. Among these, the freezing temperature scaling of the thermal component of the internal energy proposed by Rosenfeld and Tarazona~\cite{RosenfeldMolPhys1998,RosenfeldPRE2000} combines  relatively good accuracy and a relatively wide applicability~\cite{IngebrigtsenJCP2013}. In this approximation the thermal correction to the internal energy of the Yukawa fluid (the latter is dominated by the static contribution for sufficiently soft potentials, like Yukawa considered here) scales as $\propto T^{3/5}$ and hence the reduced correction is $\propto T^{-2/5}\propto \Gamma^{2/5}$. In the considered range of $\kappa$ the reduced thermal correction per particle can be well approximated as $u_{\rm th}\simeq 3.1 (\Gamma/\Gamma_{\rm fr})^{2/5}$~\cite{KhrapakJCP2015}, where $\Gamma_{\rm fr}$ is the value of the coupling parameter at freezing, which is very close to that at melting, $\Gamma_{\rm m}$ (the coexistence region is very narrow, so that usually there is no need to distinguish between $\Gamma_{\rm fr}$ and $\Gamma_{\rm m}$ for Yukawa systems~\cite{HynninenPRE2003}). The corresponding specific heat at constant volume is therefore
\begin{equation}\label{cv}
c_{\rm v}\simeq 1.5+1.86\left(\frac{\Gamma}{\Gamma_{\rm fr}}\right)^{2/5}.
\end{equation} 
The values of $\Gamma_{\rm fr}$ for various $\kappa$ have been obtained and tabulated in Ref.~\cite{HamaguchiPRE1997}  (relatively accurate fits are also available~\cite{VaulinaJETP2000,VaulinaPRE2002}.) 

The ratio $\Omega_{\rm E}/\omega_{\rm p}$ does not change much across the fluid-solid phase transition and remains approximately constant in the entire strongly coupled fluid regime~\cite{WongIEEE2018,KhrapakPoP03_2018}. We can use for example the values tabulated in Ref.~\cite{OhtaPoP2000}. All the necessary information to evaluate the equation (\ref{Horrocks}) is now in our hands. 

A comparison between the vibrational model of heat conduction and the results from MD simulations is shown in Fig.~\ref{Fig1}. The solid curves are plotted using expressions (\ref{Horrocks}) and (\ref{cv}). There is a reasonable agreement with numerical results in particular in the weak screening regime with smaller $\kappa$. For $\kappa=3$ larger deviations are observed, but in this regime the Einstein approximation is less reliable~\cite{KhrapakPoP08_2021}. Note that in the near-OCP regime with $\kappa\lesssim 1$ the thermal conductivity coefficient exhibits a very weak dependence on $\kappa$ (thus OCP result can serve as a relevant approximation in this regime). 
      
\section{Quasi-universality of the thermal conductivity coefficient}

It has been well recognized that the two other important reduced transport coefficients of strongly coupled Yukawa fluids -- the self-diffusion and shear viscosity coefficients depend quasi-universally on the relative coupling strength $\Gamma/\Gamma_{\rm fr}$. In particular, the functional form
\begin{equation}\label{Deta}
D_{\rm R} \simeq \exp\left(-3.64\sqrt{\frac{\Gamma}{\Gamma_{\rm fr}}}\right), \quad \eta_{\rm R}\simeq 0.13\exp\left(3.64\sqrt{\frac{\Gamma}{\Gamma_{\rm fr}}}\right),
\end{equation}
describes the diffusion and viscosity of strongly coupled Yukawa fluids rather well. This functional form was originally proposed for the high-density viscosity of simple neutral liquids (such as Lennard-Jones, argon, methane, and liquid sodium)~\cite{CostigliolaJCP2018}. Application to the Yukawa fluid was then discussed in Ref.~\cite{KhrapakAIPAdv2018}. The diffusion and shear viscosity coefficients in Eq.~(\ref{Deta}) are expressed in Rosenfeld's units (which is denoted by the subscript ``R''), $D_{\rm R}=D/\Delta v_{\rm T}$ and $\eta_{\rm R}=\eta\Delta^2/m v_{\rm T}$. Note that the Stokes-Einstein (SE) relation without the hydrodynamic diameter holds
\begin{equation}
D_{\rm R}\eta_{\rm R}\equiv D\eta\left(\frac{\Delta}{T}\right)= \alpha_{\rm SE},    
\end{equation}
with $\alpha_{\rm SE}\simeq 0.13$ according to Eq.~(\ref{Deta}) (but actually $\alpha_{\rm SE}$turns out to be closer to $\simeq 0.14$, according to a more accurate consideration~\cite{KhrapakPRE10_2021}). A relevant question is whether the thermal conductivity coefficient also exhibits a universal dependence on $\Gamma/\Gamma_{\rm fr}$.

The vibrational model does predict that the dependence of $\lambda_{\rm R}$ on $\Gamma/\Gamma_{\rm fr}$ be quasi-universal in the strongly coupled regime. To demonstrate this we start by rewriting Eq.~(\ref{Horrocks}) in the form
\begin{equation}\label{lR}
\lambda_{\rm R}=\frac{c_{\rm v}}{2\pi}\frac{\Omega_{\rm E}\Delta}{v_{\rm T}}=\frac{c_{\rm v}}{2\pi}\frac{\Omega_{\rm E}}{\omega_{\rm p}}\frac{\omega_{\rm p}\Delta}{v_{\rm T}}.
\end{equation}
The factor $c_{\rm v}$ is a universal function of $\Gamma/\Gamma_{\rm fr}$ as follows from Eq.~(\ref{cv}). 

Instead of using the tabulated values for $\Omega_{\rm E}/\omega_{\rm p}$ we now apply the Lindemann's melting criterion to estimate this ratio. The Lindemann melting criterion~\cite{Lindemann} states that melting of a solid
occurs when the particle root mean square vibrational amplitude
around the equilibrium position reaches a threshold value of about
$\sim 0.1$ of the interparticle distance. Its simplest variant (assuming the
Einstein model for particle vibrations in the solid state)
may be cast in the form~\cite{KhrapakPRR2020,KhrapakPoP2019}
\begin{equation}
\langle \xi^2 \rangle = \frac{3T_{\rm m}}{m\Omega_{\rm E}^2}\simeq L^2\Delta^2,
\end{equation} 
where $\langle \xi^2\rangle$ is the mean square vibrational amplitude, $T_{\rm m}\simeq T_{\rm fr}$ is the melting temperature and $L$ is the Lindemann fraction. This can be easily rewritten as
\begin{equation}\label{OmE}
\left(\frac{\Omega_{\rm E}}{\omega_{\rm p}}\right)^2=\frac{1}{\Gamma_{\rm fr}}\frac{a^2}{\Delta^2}\frac{1}{L^2}.
\end{equation}
The Lindemann's fraction is estimated by considering the one-component plasma limit ($\kappa=0$). In this limit $\Omega_{\rm E}/\omega_{\rm p}=1/\sqrt{3}$ and $\Gamma_{\rm fr}\simeq 174$~\cite{DubinRMP1999,KhrapakCPP2016}. This gives
$L\simeq 0.081$. 

Finally, the third factor in the right-hand-side of Eq.~(\ref{lR}) can be rewritten as
\begin{equation}\label{3fac}
\frac{\omega_{\rm p}\Delta}{v_{\rm T}}=\sqrt{3\Gamma}\frac{\Delta}{a}.
\end{equation} 

Combining Eqs.~(\ref{cv}), (\ref{OmE}), and (\ref{3fac}) we get
\begin{equation}\label{lambda_univ}
\lambda_{\rm R}\simeq 3.40\left[1.5+1.86\left(\frac{\Gamma}{\Gamma_{\rm fr}}\right)^{2/5}\right]\sqrt{\frac{\Gamma}{\Gamma_{\rm fr}}}.
\end{equation}  
Thus, the approximate universality is demonstrated.   
 
\begin{figure}
\includegraphics[width=8.cm]{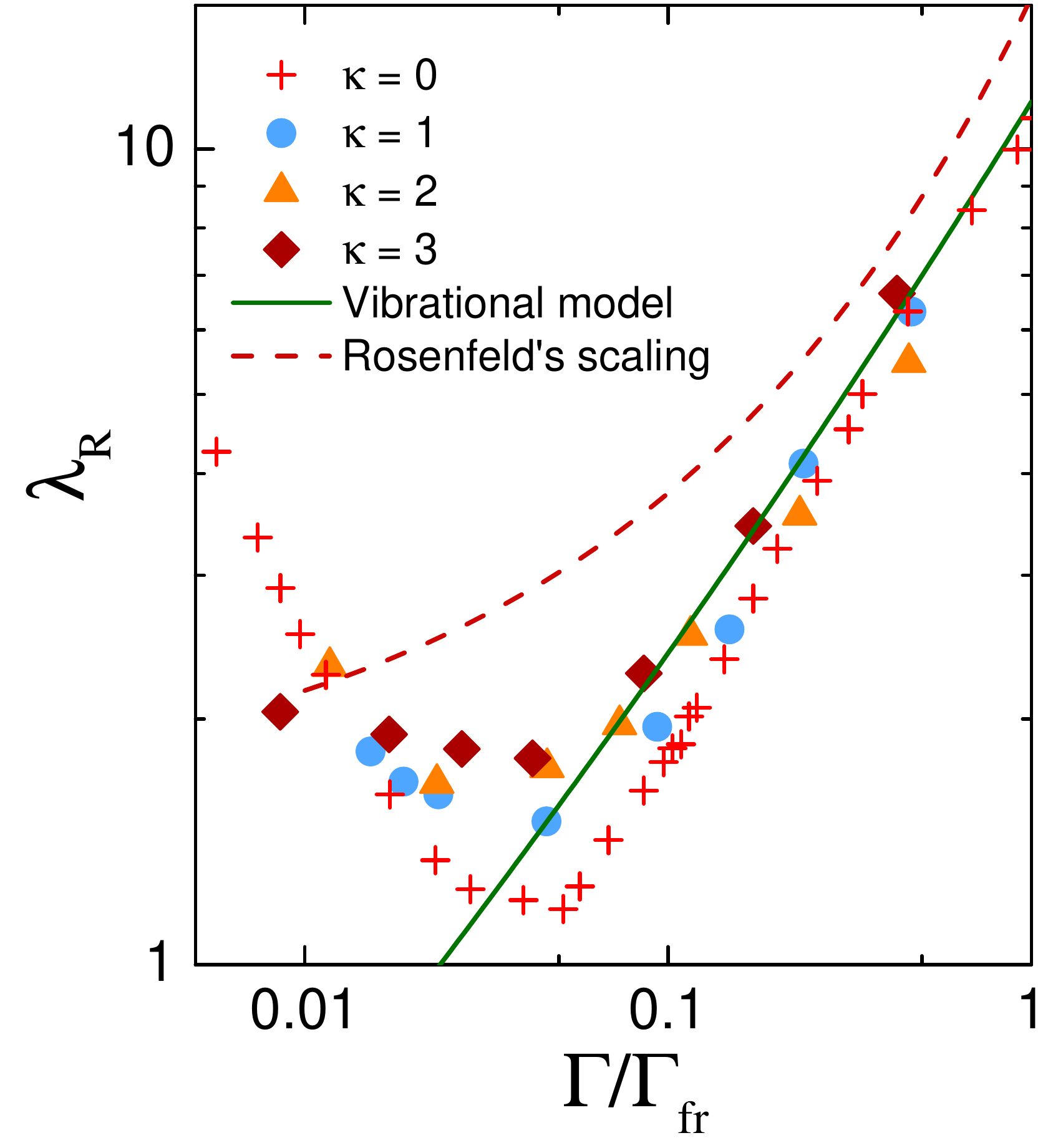}
\caption{(Color online) Reduced thermal conductivity coefficient $\lambda_{\rm R}$ versus the effective coupling parameter $\Gamma/\Gamma_{\rm fr}$. Symbols are the same as in Fig.~\ref{Fig1}. The solid curve is the result of vibrational model of Eq.~(\ref{lambda_univ}). The dashed curve corresponds to the excess entropy scaling expression from Ref.~\cite{RosenfeldJPCM1999}.} 
\label{Fig2}
\end{figure}  
 
To further illustrate this we plot the dependence of $\lambda_{\rm R}$ on $\Gamma/\Gamma_{\rm R}$ in Fig.~\ref{Fig2}. The symbols are the same as used in Fig.~\ref{Fig1}. The solid line corresponds to Eq.~(\ref{lambda_univ}).
We see that in the strongly coupled regime the MD simulation data points demonstrate some quasi-universality (although the quality of this quasi-universality is not particularly impressive). The vibrational model prediction of Eq.~(\ref{lambda_univ}) agrees reasonably well with the simulation data, taking into account their scattering. The dashed curve corresponds to the excess entropy scaling of the form $\lambda_{\rm R}\simeq 1.5 \exp(-0.5s_{\rm ex})$~\cite{RosenfeldJPCM1999}, where the excess entropy is approximated as $s_{\rm ex}\simeq -4.65(\Gamma/\Gamma_{\rm fr})^{2/5}$. We see that this formula needs improvement in the case of the strongly coupled Yukawa fluid. 

Let us further elaborate on some general trends exhibiting by the thermal conductivity coefficient. The observed quantitative behaviour is quite general and resembles that in other simple fluids such as the Lennard-Jones and hard spheres fluids, liquefied noble gases, and other atomic and molecular fluids~\cite{TrachenkoSciAdv2020,TrachenkoPRB2021,TrachenkoPhysToday2021,
KhrapakPoF2022,KhrapakJCP2022_1}. The emergence of the minimum of $\lambda_{\rm R}$ is related to the crossover between gas-like dynamics at low densities and liquid-like dynamics at high densities. The idea that the kinematic viscosity and thermal diffusivity of liquids and supercritical fluids have lower bounds, which are determined by fundamental physical constants has been recently discussed~\cite{TrachenkoSciAdv2020,TrachenkoPRB2021,TrachenkoPhysToday2021}. It turns out, however, that purely classical arguments suffice to understand the magnitude of the minimum shear viscosity and thermal conductivity coefficients at their respective minima as well as the location of these minima~\cite{KhrapakPoF2022}. This topic is of quite general interest and hence let us discuss it in some further detail.

The thermal conductivity in the dilute gas regime is determined by pairwise collisions between the particles. Elementary kinetic formula for the thermal conductivity coefficient yields~\cite{LifshitzKinetics} $\lambda\sim c_{\rm p}v_{\rm T}\rho\ell$, where $\ell$ is the mean free path between collisions. It is related to the momentum transfer cross section $\sigma$ via $\ell=1/\rho \sigma$. In liquids, on the other hand, dynamics is mainly controlled by collective effects. A
natural transition condition between the gas-like and liquid-like dynamical regimes can be defined as the point where the effective momentum transfer cross section becomes comparable to the interatomic separation squared, $\sigma \sim \Delta^2$. This condition was in fact previously used to discriminate between the “ideal” (gas-like) and “nonideal” (fluid-like) regions on the phase diagram of Yukawa systems (complex plasmas)~\cite{KhrapakPRE2004_MT}. If we extrapolate the kinetic formula to the transition point, then we obtain $\lambda\sim c_{\rm p}v_{\rm T}/\sigma\sim c_{\rm p}v_{\rm T}/\Delta^2$ and $\lambda_{\rm R}\sim c_{\rm p}$. Assuming that the specific heat at the transition point is not much different from its ideal gas value, we arrive at $\lambda_{\rm R}^{\rm min}\sim \tfrac{5}{2}$ for mono-atomic liquids (for molecular liquids the minimum value should be higher due to the presence of additional degrees of freedom). The actual values of $\lambda_{\rm R}$ at the minimum in Fig.~\ref{Fig2} are slightly above unity for the one-component plasma limit and approach $\lambda_{\rm R}^{\rm min}\simeq 2$ for $\kappa=3$. This indicates that in the regime of soft interactions the value of $\lambda_{\rm R}^{\rm min}$ is not truly universal and somewhat increases with increasing the potential steepness. For steeper interactions saturation occurs at $\lambda_{\rm R}^{\rm min}\simeq 3$, as observed for the HS and LJ fluids and liquefied noble gases~\cite{KhrapakPoF2022}.

The location of the minimum can be estimated as follows. The transition between the gas-like and liquid-like regimes correspond to the condition that the characteristic time between pairwise collisions becomes comparable with the inverse oscillation frequency due to collective interactions. This implies that $\Omega_{\rm E}/2\pi\sim v_{\rm T}/\ell\sim v_{\rm T}/\Delta$ at the transition point. The following condition of the crossover emerges
\begin{equation}
\frac{\Omega_{\rm E}\Delta}{v_{\rm T}}\sim \frac{\sqrt{3}}{L}\sqrt{\frac{\Gamma}{\Gamma_{\rm fr}}}\sim 2\pi.
\end{equation}
The location of the minimum should hence be expected at 
$\Gamma/\Gamma_{\rm fr}\sim 0.1$, in reasonable agreement with what is observed in Fig.~\ref{Fig2}. Note that the minimum in the reduced shear viscosity also occurs at  $\Gamma/\Gamma_{\rm fr}\sim 0.1$~\cite{KhrapakAIPAdv2018}. This should be expected, because both minima are related with crossover between gas-like and liquid-like regimes.   

According to Eq.~(\ref{lambda_univ}) the value of $\lambda_{\rm R}$ at the freezing point is $\lambda_{\rm R}\simeq 11$. This is very close to $\lambda_{\rm R}\sim 10$ of various models as well as real atomic and molecular liquids: Lennard-Jones, Ne, Ar, Kr, Xe, N$_2$, O$_2$, CO$_2$, and CH$_{4}$~\cite{KhrapakPoF2022,KhrapakJETPLett2021,KhrapakJCP2022_1}. This potentially useful universality is violated for the HS fluid, which has $\lambda_{\rm R}\simeq 14$ at the freezing point~\cite{Pieprzyk2020,HSNew}.     

\section{Two-dimensional complex plasma layers}

The problem of transport coefficients in 2D systems has been elusive for quite a long time. The absence of valid transport coefficients in 2D systems was predicted based on the analysis of the velocity autocorrelation function and of the kinetic
parts of the correlation functions for the shear viscosity and the heat conductivity~\cite{ErnstPRL1970}. Despite this general conclusion, a number of contradicting examples have been presented over the years. For example, molecular-dynamics simulations of the 2D classical electron system (2D Coulomb fluid) yielded indications for the existence of a self-diffusion coefficient~\cite{HansenPRL1979}. Strong indications of normal self-diffusion in 2D Yukawa fluids at sufficiently long time scales were also reported~\cite{OttPRL2009}. Existence of finite shear viscosity coefficients of strongly coupled 2D Yukawa fluids was demonstrated in experiments with complex (dusty) plasma monolayers and molecular dynamics (MD) simulations~\cite{DonkoMPLB2007}. In a dedicated study of Ref.~\cite{DonkoPRE2009} it was observed that the  self-diffusion coefficient exists at sufficiently high temperatures, the viscosity coefficient exists at sufficiently low temperatures, but not in the opposite limits. The thermal conductivity coefficient does not appear to exist at high temperature. For low temperatures no definite conclusion could be drawn, because of the technical challenges posed by signal-to-noise ratios and a long initial decay of the corresponding correlation function~\cite{DonkoPRE2009}.

An outstanding review of thermal conduction in classical low-dimensional lattices can be found in Ref.~\cite{LepriPR2003}. The mode-coupling theory predicts finite thermal conductivity coefficient in 3D and its divergence in the thermodynamic limit for lower dimensions: $\lambda\propto \ln N$ (in 2D) and $\lambda\propto N^{2/5}$ (in 1D), where $N$ is the number of particles in the system. Nevertheless, numerous indications in favor of the finiteness of $\lambda$ in the thermodynamic limit of low dimensional systems have also been reported over the years~\cite{LepriPR2003}.

Considering 2D complex (dusty) plasma monolayers, finite values of the thermal conductivity coefficient were measured in crystalline, melted, and fluid complex (dusty) plasmas~\cite{NunomuraPRL2005,FortovPRE2007,NosenkoPRL2008,
GoreePPCF2013,DuPRE2014}.
Numerical simulations for conditions representative to those in laboratory experiments also yelded finite values~\cite{HouJPA2009,KhrustalyovPRE2012,ShahzadPoP2015}.

A relevant question is: How these reported results would compare with the prediction of the vibrational mechanism of heat transfer? This question was addressed in Ref.~\cite{KhrapakPoP2021}. We provide a brief summary below.

First of all, the vibrational model should be generalized to the 2D regime. This is straightforward~\cite{KhrapakPoP2021}, and the result is
\begin{equation}\label{Cond1}
\lambda=c_{\rm v}\frac{\langle \omega \rangle}{2\pi}=c_{\rm v}\frac{\Omega_{\rm E}}{2\pi},
\end{equation}     
where in the last equation the Einstein approximation for atomic vibrations is used. As a more accurate approximation, averaging over the long-wavelength spectrum containing one longitudinal and one transverse acoustic-like modes with acoustic dispersions can be performed. For details see Ref.~\cite{KhrapakPoP2021}. 

A comparison was then performed with some available results from experiments and simulations. In particular, the experiment with 2D complex plasma undergoing a solid-fluid phase transition and performed by Nosenko {\it et al.}~\cite{NosenkoPRL2008} was used. A non-equilibrium Brownian dynamics simulation by Hou and Piel~\cite{HouJPA2009} and an equilibrium Langevin molecular dynamics simulation by Khrustalyov and Vaulina~\cite{KhrustalyovPRE2012} 
were also selected for comparison. Additionally, a homogenous nonequilibrium molecular dynamics simulation by Shahzad and He~\cite{ShahzadPoP2015} was considered, but the comparison was not conclusive in this case. 

Overall, it was reported that for 2D complex (dusty) plasma layers the vibrational model of heat transfer is not inconsistent with the results from experiments and simulations for which comparison is possible and meaningful. Generally, the model is able to reproduce reasonably well the numerical values of the heat conductivity coefficient. On the other hand, it was pointed out that more results are needed to get more confidence regarding its applicability conditions. In particular, the dependence of the thermal conductivity coefficient on the system size was not yet investigated systematically.   

\section{Conclusion}

The main conclusion can be formulated as follows. The vibrational model of heat transfer works reasonably well for the strongly coupled Yukawa fluid. This model relates the coefficient of thermal conductivity to the specific heat, mean inter-atomic separation and an average vibrational frequency of atoms around their temporary equilibrium positions. For soft plasma-related interactions an assumption that all atoms vibrate with the same Einstein frequency (Einstein approximation) is appropriate. The model demonstrates a rather good agreement with the results from numerical simulations in both three-dimensional and two-dimensional geometries, although in two-dimensional case some questions still remain.  


\bibliography{TC_Ref}

\end{document}